\documentclass[twocolumn,aps,superscriptaddress,showpacs,nofootinbib,floatfix]{revtex4}
\usepackage{bm,feynmf}
\usepackage{amsmath}
\usepackage{graphicx}
\usepackage{lmodern}
\usepackage{amsmath}
\usepackage{amssymb}
\usepackage{amsfonts}
\usepackage{float}
\usepackage[T1]{fontenc}
\usepackage[utf8]{inputenc}
\usepackage[normalem]{ulem}  
\usepackage{xcolor}

\renewcommand\sout{\bgroup \color{red} \ULdepth=-.5ex \ULset}
\begin{document}


\title{Heavy quark diffusion from coherent color fields in relativistic heavy-ion collisions}


\author{Taesoo Song}\email{song@fias.uni-frankfurt.de}
\affiliation{Frankfurt Institute for Advanced Studies and Institute for Theoretical Physics, Johann Wolfgang Goethe Universit\"{a}t, Frankfurt am Main, Germany}
\author{Thomas Epelbaum}\email{epelbaum@physics.mcgill.ca}
\affiliation{McGill University, Department of Physics 3600 rue University, Montr\'{e}al QC H3A 2T8, Canada}


\begin{abstract}
The diffusion coefficients of heavy quarks from the coherent color electromagnetic fields which are generated in the early stage of relativistic heavy-ion collisions are calculated at midrapidity, and compared with those obtained from collisions within a thermalized quark-gluon plasma.
The coherent color fields are modeled such that they are initially longitudinal and then become isotropic.
We found that the diffusion coefficients from the coherent color fields are larger than those from collisions except for very fast heavy quarks, and the color fields are less effective for heavy-quark energy loss.
The importance of coherent color fields for heavy-quark diffusion decreases as energy density decreases.
\end{abstract}


\maketitle

\section{introduction}

The color glass condensate (CGC) is one of the promising models meant to describe the initial state of relativistic heavy-ion collisions.
The model is so named, because the nuclei participating in ultra-relativistic heavy-ion collisions are composed of color fields which slowly evolve with time and are highly occupied.

As the nucleus energy increases, the number of gluons in the nucleus rapidly increases while the cross section for the nucleus scattering slowly increases.
This implies that the gluon fields in the nucleus are highly occupied at high energy, and should be saturated by nonlinear interactions.
Due to the high density, the distance between gluons is so short that the strong coupling strength is in fact small.
However, the gluon interactions are coherent and become strong as the gravitational interactions do~\cite{McLerran:2010uc}.

In extremely high energy collisions, the nucleus is like an infinitely thin sheet of color glass.
Partons with large energy fractions in the nucleus play the role of color sources, and those with small energy fractions are described as the classical color fields generated by the sources.
The color sources are distributed on the color sheet with the transverse size of the order of the inverse saturation momentum, and the color electromagnetic fields have only transverse components before heavy-ion collision.

Rapidity dependence ignored, one can numerically calculate color fields in space-time by solving the classical Yang-Mills equations.
At the contact time of two color sheets, longitudinal color electromagnetic fields are instantaneously generated in forward light cone, and then transverse color electromagnetic fields grow till they are comparable to the longitudinal ones~\cite{Chen:2013ksa,Lappi:2006fp}.
One cannot reach isotropic color electromagnetic fields only by solving the classical Yang-Mills equations, and quantum corrections seem to be needed~\cite{Gelis:2013rba}.
The nuclear matter between the two color glasses before the formation of an isotropic quark-gluon plasma (QGP) is called the glasma.

Heavy flavor is one of the important probes in search of the properties of the hot dense nuclear matter created in relativistic heavy-ion collisions.
It has been found that the nuclear modification factor and elliptic flow of heavy flavors are not small in relativistic heavy-ion collisions, which indicates strong interactions of heavy flavors with the hot dense nuclear matter~\cite{Adamczyk:2014uip,Tlusty:2012ix,ALICE:2012ab,Abelev:2013lca}.
There are numerous theoretical studies to explain experimental data, where heavy flavor interacts with nuclear matter through collision and gluon radiation~\cite{GolamMustafa:1997id,Moore:2004tg,Zhang:2005ni,Molnar:2006ci,vanHees:2005wb,
Gossiaux:2010yx,
Gossiaux:2012ya,Ozvenchuk:2014rpa,Cao:2013ita,Alberico:2011zy,Sharma:2009hn,He:2011qa,
Akamatsu:2008ge,BAMPS,Lang:2012yf,Berrehrah:2014kba,Berrehrah:2014tva,Berrehrah:2015ywa,
Das:2013kea,Das:2015ana,Song:2015sfa}.

In the CGC picture, relativistic heavy-ion collision generates strong coherent color fields between two receding heavy nuclei. Since heavy flavor is early produced, it will interact with the strong color fields. The production time of heavy quark pair is at most $1/(2m_Q)$ with $m_Q$ being the heavy quark mass.
Considering the masses of charm and bottom quarks being respectively about 1.5 and 5.0 GeV, the production times are less than 0.07 and 0.02 $\rm fm/c$.
They are smaller than the typical time scale during which the CGC model should be valid, especially for the bottom quark. The heavy quark interactions with the strong coherent color fields take place mainly before initial thermalization while collisions or gluon radiations of heavy quarks in QGP, which have been widely studied, take place after that.
Since the coherent color fields have a random color and a random direction, the heavy quarks traveling through these fields will diffuse as in a thermalized QGP.

In this study, we calculate the diffusion coefficients of a heavy quark from coherent color fields in relativistic heavy-ion collisions and compare them with those obtained from collisions within a thermalized QGP.

This paper is organized as follows:
The diffusion coefficients of heavy flavor induced by color fields are calculated in Sec.~\ref{coherent}, and they are computed for realistic relativistic heavy-ion collisions in Sec.~\ref{cgc}.
In Sec.~\ref{collision}, these diffusion coefficients are then compared with those obtained from collisions within a thermalized QGP, and a summary is given in Sec.~\ref{summary}.

\section{Diffusion coefficients in color fields}\label{coherent}

Let us start with the nonrelativistic Langevin equations expressed as~\cite{Moore:2004tg}
\begin{eqnarray}
\frac{dp_L}{dt}&=&-\eta_D(p)p+\xi_L,\nonumber\\
\frac{dp_T}{dt}&=&\xi_T,
\label{langevin}
\end{eqnarray}
where $p_L$ and $p_T$ are, respectively, the longitudinal and transverse momenta of a particle, $\eta_D$ a momentum drag coefficient, and $\xi_L$ and $\xi_T$ random momentum kicks in the longitudinal and transverse directions which satisfy

\begin{eqnarray}
\langle \xi_L^i(t)\xi_L^j(t^\prime)\rangle&=&\kappa_L(p)\hat{p}^i\hat{p}^j \delta(t-t^\prime),\nonumber\\
\langle \xi_T^i(t)\xi_T^j(t^\prime)\rangle&=&\kappa_T(p)(\delta^{ij}-\hat{p}^i\hat{p}^j) \delta(t-t^\prime),\nonumber\\
\langle \xi_T^i(t)\xi_L^j(t^\prime)\rangle&=&0,
\label{random}
\end{eqnarray}
where $\langle \ldots \rangle$ represents ensemble average, $\kappa_L(p)$ and $\kappa_T(p)$ are the mean-squared longitudinal and transverse momentum transfers per unit time, and $\hat{\bf p}$ is the unit vector of the particle three-momentum.
$\delta(t-t^\prime)$ in Eq.~(\ref{random}) implies that the random forces at two different times are completely uncorrelated.
However, it is not true in reality, because any force acting on a particle cannot be as sharp as a delta function.
One may use a normalized gaussian function with a finite width instead of a delta function. This is the strategy we will adopt in the following.
Integrating Eq.~(\ref{random}) over $t^\prime$ leads to
\begin{eqnarray}
\kappa_L(p)&=&2\hat{p}_i\hat{p}_jD^{ij},\nonumber\\
\kappa_T(p)&=&(\delta_{ij}-\hat{p}_i\hat{p}_j)D^{ij},
\label{kappa0}
\end{eqnarray}
where $D_{ij}$ is a diffusion tensor defined as~\cite{Majumder:2006wi}
\begin{eqnarray}
D^{ij}=\frac{1}{2}\int^\infty_{-\infty}dt^\prime \langle \xi^i(t^\prime)\xi^j(t)\rangle=\int^t_{-\infty}dt^\prime \langle \xi^i(t^\prime)\xi^j(t)\rangle,
\label{dij}
\end{eqnarray}
where $\xi^i=\xi_L^i+\xi_T^i$ and we have assumed that the correlator is a function of $|t-t'|$,
an assumption that makes  sense for a thermal bath at constant temperature. We note that the random forces in Eq.~{(\ref{dij}) do not need the subscript $L$ or $T$ by virtue of the tensor structures in Eq.~(\ref{kappa0}).

\begin{figure}[h]
\centering
\includegraphics[width=0.45\textwidth]{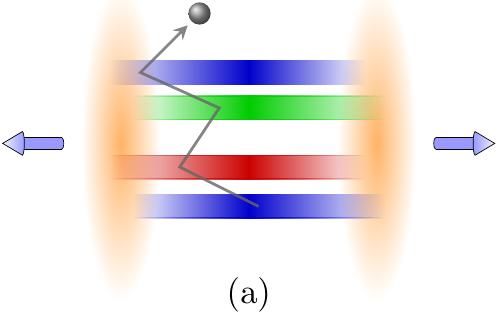}
\includegraphics[width=0.45\textwidth]{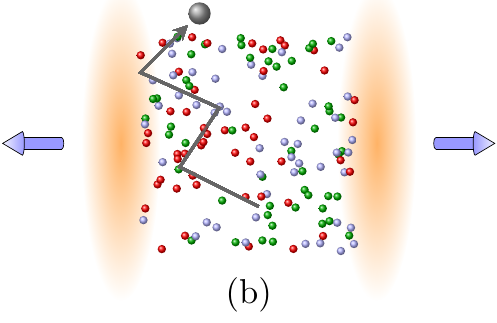}
\caption{(Color online) Schematic figures for a heavy quark scattering with initial coherent color fields (a) and within a thermalized QGP (b) in relativistic heavy-ion collisions.}
\label{fig1}
\end{figure}

Fig.~\ref{fig1} shows schematic cartoons for a heavy quark scattering with initial coherent color fields and within a thermalized QGP in relativistic heavy-ion collisions.
Since the initial coherent fields have a random color and a random leftward or rightward direction, with a typical transverse size of the order of the inverse saturation momentum, a heavy quark interaction with the fields is similar with random scatterings in a thermalized QGP.
Therefore, the random force in Eq.~(\ref{dij}) can be substituted by the color Lorentz force, ${\bf F}=g_s Q^a({\bf E}^a+{\bf v}\times {\bf B}^a)$ where $g_s$, $Q^a$ and ${\bf v}$ are respectively the strong coupling constant, the charge of color $a$, and the velocity of a parton in the color electric and magnetic fields, ${\bf E}^a$ and ${\bf B}^a$.
The correlation of the color Lorentz force is then expressed as
\begin{eqnarray}
\langle F_i(x^\prime)F_j(x)\rangle=g_s^2\frac{C_2\delta^{ab}}{N_c^2-1}\bigg[\langle E_i^a(x^\prime)E_j^b(x)\rangle\nonumber\\
+\varepsilon_{ikl}\varepsilon_{jmn}v_kv_m\langle B^a_l(x^\prime)B^b_n(x)\rangle\bigg],
\end{eqnarray}
where $N_c$ is the number of colors, $C_2=1/2$ for quark and $C_2=N_c$ for gluon from $Q^aQ^b=C_2\delta^{ab}/(N_c^2-1)$, $\varepsilon_{ikl}$ and $\varepsilon_{jmn}$ are the usual Levi-Civita antisymmetric tensors, and $\langle E_i B_j\rangle=\langle B_i E_j\rangle=0$ is assumed~\cite{Asakawa:2006jn,Litim:2001db}.
Identifying the color Lorentz force with the medium kicks in Eq. (\ref{dij}) leads to the following expressions for $\kappa_L$ and $\kappa_T$
\begin{eqnarray}
\kappa_L=\frac{2g_s^2 C_2}{N_c^2-1}\int^t_{-\infty}dt^\prime \bigg\langle E_L^a(x^\prime) E_L^a(x)\bigg\rangle,~~~~~~~~~~~~~\nonumber\\
\kappa_T=\frac{g_s^2 C_2}{N_c^2-1}\int^t_{-\infty}dt^\prime\bigg[\bigg\langle \vec{E}^a(x^\prime)\cdot\vec{E}^a(x)\bigg\rangle~~~~~~~~~~~ \nonumber\\
-\bigg\langle E_L^a(x^\prime) E_L^a(x)\bigg\rangle+\bigg\langle \vec{v}(x^\prime)\cdot\vec{v}(x)\vec{B}^a(x^\prime)\cdot\vec{B}^a(x)\bigg\rangle \nonumber\\
-\bigg\langle\vec{v}\cdot \vec{B}^a(x^\prime) \vec{v}\cdot \vec{B}^a(x)\bigg\rangle\bigg],
\label{kappa}
\end{eqnarray}
where $E_L\equiv \hat{p}\cdot \vec{E}^a$ is defined as the longitudinal color electric field. As mentioned below Eq.~(\ref{random}), random forces are spread with finite widths in space-time.
Therefore, we introduce a Gaussian form for the correlation of color electromagnetic fields at two different space-time:
\begin{eqnarray}
\langle E_i^a(x^\prime)E_j^a(x)\rangle=\delta_{ij}\langle E_i^aE_i^a\rangle(x) \exp\bigg[-\sum_\mu\frac{(x_\mu-x_\mu^\prime)^2}{2\sigma^2_{\mu}}~\bigg],\nonumber\\
\langle B_i^a(x^\prime)B_j^a(x)\rangle=\delta_{ij}\langle B_i^aB_i^a\rangle(x) \exp\bigg[-\sum_\mu\frac{(x_\mu-x_\mu^\prime)^2}{2\sigma^2_{\mu}}~\bigg],
\label{correlation}
\end{eqnarray}
where $\sigma_{\mu}$ is the correlation length of color electromagnetic fields in $\mu$ direction.

Since we study a heavy quark in coherent color fields, $C_2=1/2$, and the velocity does not change much: $\vec{v}(x^\prime)\approx\vec{v}(x)\equiv \vec{v}$.
Substituting Eq.~(\ref{correlation}) into Eq.~(\ref{kappa}), $\kappa_L$ and $\kappa_T$ are simplified into
\begin{eqnarray}
\kappa_L&=&\frac{g_s^2\tau_m}{N_c^2-1} \langle (E_L^a)^2\rangle,\nonumber\\
\kappa_T&=&\frac{g_s^2\tau_m}{2(N_c^2-1)} \langle (E_T^a)^2+ v^2(B_T^a)^2\rangle,
\label{kappa2}
\end{eqnarray}
where $E_T$ and $B_T$ are respectively the transverse color electric and magnetic fields, and $\tau_m$ is the memory time of the color electromagnetic field defined as
\begin{eqnarray}
\tau_m\equiv \int_{-\infty}^t dt^\prime\exp\bigg[-\sum_\mu\frac{(x_\mu-x_\mu^\prime)^2}{2\sigma^2_{\mu}}~\bigg]~~~~~~~~~~~~~~\nonumber\\
=\int_{-\infty}^t dt^\prime\exp\bigg[-\frac{(t-t^\prime)^2}{2\sigma_0^2}-\sum_{i=1,2,3}\frac{v_i^2(t-t^\prime)^2}{2\sigma_i^2}~\bigg].
\label{memoryt}
\end{eqnarray}
It is interesting to notice that since magnetic fields induce only transverse force, they do not contribute to $\kappa_L$ in Eq.~(\ref{kappa2}).

For a mid-rapidity heavy quark in relativistic heavy-ion collisions, the memory time is given by
\begin{eqnarray}
\tau_m&=&\frac{1}{2}\int_{-\infty}^\infty dt^\prime\exp\bigg[-\frac{(t-t^\prime)^2}{2}\bigg\{\frac{1}{\sigma_0^2}+\frac{v_T^2}{\sigma_T^2}\bigg\}~\bigg]\nonumber\\
&=&\sqrt{\frac{\pi}{2}}\bigg[\frac{1}{\sigma_0^2}+\frac{v_T^2}{\sigma_T^2}\bigg]^{-1/2},
\end{eqnarray}
where $v_T$ and $\sigma_T$ are, respectively, the transverse velocity of a heavy quark and the transverse correlation length of color electromagnetic fields.
In the CGC, the latter is approximately the inverse of saturation momentum, $\sigma_T\simeq 1/Q_s$~\cite{McLerran:2010uc}, which is around 0.1$\sim$0.2 ${\rm fm}$ at the Relativistic Heavy Ion Collider (RHIC) and the Large Hadron Collider (LHC) energies~\cite{Iancu:2003xm}.
Assuming boost invariance, $\sigma_0$ is roughly the lifetime of coherent color fields and expected to be larger than $\sigma_T$.
For a fast heavy quark ($v_T \gg \sigma_T/\sigma_0$), the memory time approximates to
\begin{eqnarray}
\tau_m \simeq \sqrt{\frac{\pi}{2}}(Q_s v_T)^{-1},
\label{approx1}
\end{eqnarray}
and for a slow heavy quark ($v_T \ll \sigma_T/\sigma_0$),
\begin{eqnarray}
\tau_m \simeq \sqrt{\frac{\pi}{2}}\sigma_0.
\label{approx2}
\end{eqnarray}

From the Fixed-Order Next-to-Leading Logarithm (FONLL) calculations~\cite{Cacciari:2005rk}, the average transverse momenta of mid-rapidity charm quarks are respectively 1.5 $\rm GeV$ and 2.4 $\rm GeV$ in p+p collisions at $\sqrt{s_{\rm NN}}=200~{\rm GeV}$ and $2.76~{\rm TeV}$, which correspond to $v_T=$0.7 and 0.85.
For mid-rapidity bottom quarks, the average transverse velocities are respectively 0.58 and 0.7.
Therefore, we take the approximation of Eq.~(\ref{approx1}) in this study.

\section{Application to relativistic heavy-ion collisions}\label{cgc}

The CGC model introduces a separation scale $x_0$: the partons which have larger energy fractions than $x_0$ are treated as static color sources while those with smaller energy fractions as classical color fields generated by the sources.
The color sources are located near the color sheets, which are the Lorentz-contracted nuclei, with a transverse size of the order of the inverse saturation momentum.
After two color sheets pass through each other in relativistic heavy-ion collisions, the classical color fields are calculated in forward light cone by numerically solving the classical Yang-Mills equations, satisfying the boundary conditions given by color sources on the light cone~\cite{Chen:2013ksa,Lappi:2006fp}.
The numerical calculations assume boost invariance, which means that the physics does not depend on $\eta \sim \ln[(t+z)/(t-z)]$.
Though it seems similar to the capacitor problem in electrodynamics, color sources are randomly distributed on the sheet with a characteristic transverse size, and nonabelian gauge equations are to be solved.
The results show that longitudinal color electric and magnetic fields are instantaneously generated at the contact time of two color sheets, and then they diminish with time.
On the other hand, transverse color electric and magnetic fields, which are initially absent in forward light cone, grow with time till they are comparable to longitudinal ones in strength~\cite{Chen:2013ksa,Lappi:2006fp}.
In other words, the initial color electromagnetic fields are given by
\begin{eqnarray}
(E_z^a)^2\simeq (B_z^a)^2\neq 0,\nonumber\\
E_{x,y}^a=B_{x,y}^a=0,~~
\label{initial}
\end{eqnarray}
and the asymptotic final ones by
\begin{eqnarray}
(E_z^a)^2\simeq (E_x^a)^2+(E_y^a)^2,\nonumber\\
(B_z^a)^2\simeq (B_x^a)^2+(B_y^a)^2,
\label{final}
\end{eqnarray}
and more details can be found in Appendix B. From the energy-momentum tensor of pure gauge Quantum Chromodynamics (QCD) Lagrangian,
\begin{eqnarray}
T_{\mu\nu}=\frac{1}{4}g_{\mu\nu}F^{\alpha\beta a}F_{\alpha\beta}^a-g^{\alpha\beta}F^{a}_{\mu\alpha} F_{\nu\beta}^a,
\end{eqnarray}
energy density and pressure are, respectively, given by
\begin{eqnarray}
\varepsilon=T_{00}=\frac{1}{2}\bigg\{(E^a)^2+(B^a)^2\bigg\},\nonumber\\
p_i=T_{ii}=\varepsilon-E_i^{a2}-B_i^{a2},~~~~~~~\label{pressure}
\end{eqnarray}
where $i=1 \sim 3$,$E^a_i=F^a_{0i}$ and $B^a_i=-(1/2)\epsilon_{ijk}F^{ajk}$.
Substituting Eq.~(\ref{initial}) and (\ref{final}) into Eq.~(\ref{pressure}), initial pressures are given by
\begin{eqnarray}
p_x=p_y=\varepsilon,~~~p_z=-\varepsilon,
\label{initialp}
\end{eqnarray}
and final ones by
\begin{eqnarray}
p_x=p_y=\varepsilon/2,~~~p_z=0.
\label{finalp}
\end{eqnarray}

Since the initial color fields are expected to eventually turn into isotropic gluon gas, it seems that the classical Yang-Mills equations are not sufficient, and quantum corrections are needed in order for the complete time-evolution of initial gluonic matter to show signs of isotropization~\cite{Gelis:2013rba}.

In order to deal with the anisotropic pressure, we introduce a parameter defined by $\zeta\equiv p_z/p_x\simeq p_z/p_y$.
The squared color electric fields are then expressed as

\begin{eqnarray}
(E^a_x)^2+(E^a_y)^2&=&\frac{\zeta+1}{\zeta+2}(E^a)^2,\nonumber\\
(E^a_z)^2&=&\frac{1}{\zeta+2}(E^a)^2,
\label{aniso1}
\end{eqnarray}
and same for color magnetic fields.
We note that $\zeta=-1$ at the contact time of two color sheets, and $\zeta$ should evolve towards $+1$, which corresponds to isotropic color fields. In practice, $\zeta$ converges towards $0$ (free-streaming) for the late time behavior of classical Yang-Mills evolution with classical initial conditions.

From Eq.~(\ref{aniso1}), the longitudinal and transverse components of color electromagnetic fields
for a mid-rapidity heavy quark which moves in transverse $(x,y)$ direction are respectively given by

\begin{eqnarray}
\langle (E_L^a)^2 \rangle=\frac{\zeta+1}{2(\zeta+2)}\langle(E^a)^2\rangle,\nonumber\\
\langle (E_T^a)^2 \rangle=\frac{\zeta+3}{2(\zeta+2)}\langle(E^a)^2\rangle,
\label{aniso2}
\end{eqnarray}

and the same equations for color magnetic fields.
Substituting Eq.~(\ref{approx1}) and (\ref{aniso2}) into Eq.~(\ref{kappa2}),
\begin{eqnarray}
\kappa_L&=&\frac{\sqrt{2}\pi^{3/2}\alpha_s}{(N_c^2-1)Q_s}\frac{\zeta+1}{\zeta+2} \bigg\langle\frac{(E^a)^2}{v}\bigg\rangle,\nonumber\\
\kappa_T&=&\frac{\pi^{3/2}\alpha_s}{\sqrt{2}(N_c^2-1)Q_s}\frac{\zeta+3}{\zeta+2} \bigg\langle\frac{(E^a)^2}{v}+v(B^a)^2\bigg\rangle,
\label{kappa3}
\end{eqnarray}
where $\alpha_s=g_s^2/(4\pi)$. From the relation $(E^a)^2= (B^a)^2= \varepsilon$ derived in appendix A, Eq.~(\ref{kappa3}) is finally expressed in term of the energy density as:
\begin{eqnarray}
\kappa_L&=&\frac{\sqrt{2}\pi^{3/2}\alpha_s\varepsilon}{(N_c^2-1)Q_s}\frac{\zeta+1}{\zeta+2}v^{-1},\nonumber\\
\kappa_T&=&\frac{\pi^{3/2}\alpha_s\varepsilon}{\sqrt{2}(N_c^2-1)Q_s}\frac{\zeta+3}{\zeta+2} \bigg(\frac{1}{v}+v\bigg).
\label{kappaf}
\end{eqnarray}

It seems that Eq.~(\ref{kappaf}) diverges for a static heavy quark.
In this case, however, one should use Eq.~(\ref{approx2}) for memory time instead of Eq.~(\ref{approx1}).

\begin{figure}[h]
\centering
\includegraphics[width=0.5\textwidth]{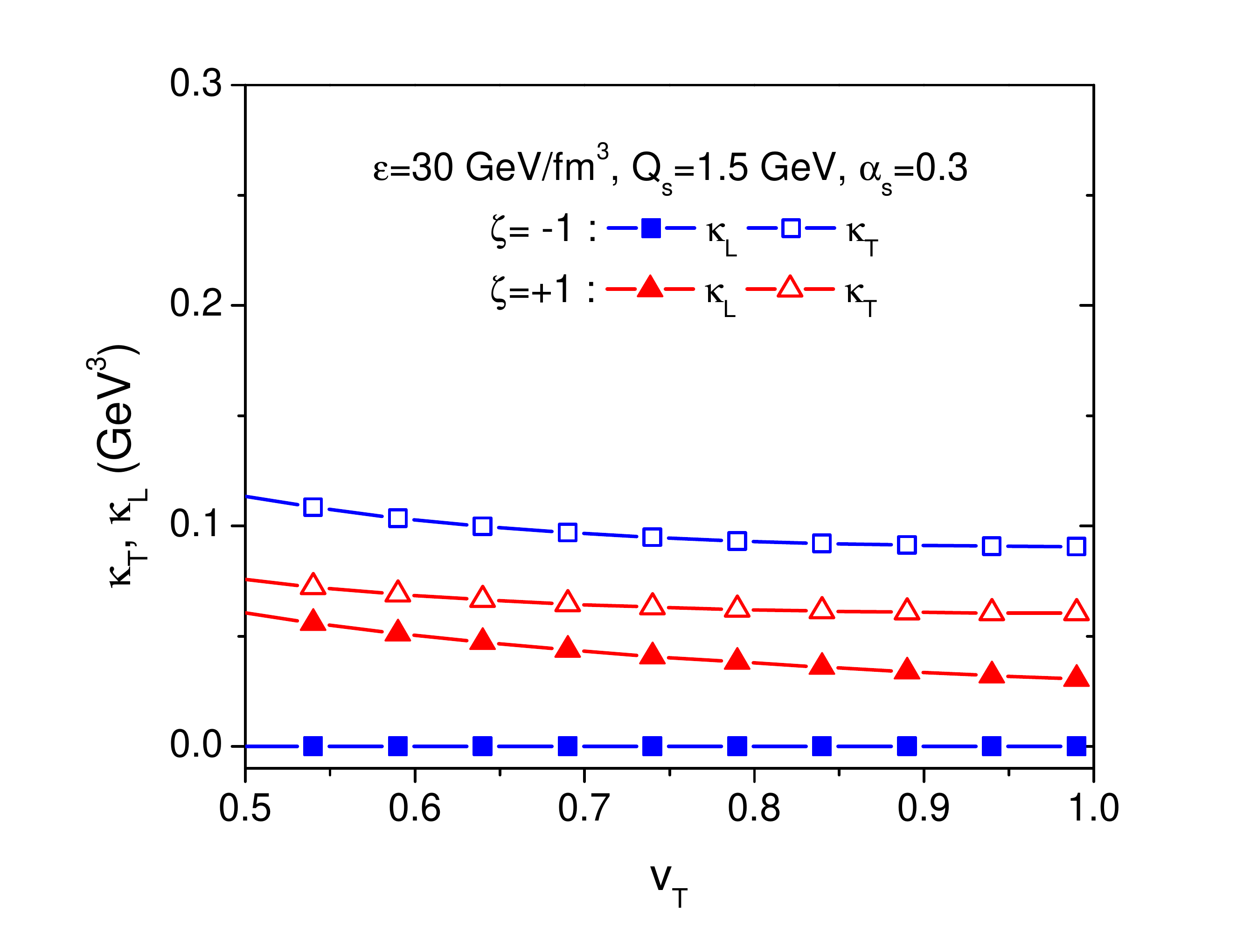}
\caption{(Color online) $\kappa_L$ (filled) and $\kappa_T$ (open) of a mid-rapidity heavy quark from the initial (blue) and isotropic (red) coherent color fields, which correspond to $\zeta=-1$ and $\zeta=+1$ respectively. Energy density, the saturation momentum, and $\alpha_s$ are, respectively, taken to be $30~{\rm GeV/fm^3}$, $1.5~{\rm GeV}$, and 0.3.}
\label{example}
\end{figure}

Fig.~\ref{example} shows $\kappa_L$ and $\kappa_T$ of a mid-rapidity heavy quark from initial and isotropic coherent color fields.
The energy density, the saturation momentum, and $\alpha_s$ are respectively taken to be $30~{\rm GeV/fm^3}$, $1.5~{\rm GeV}$, and 0.3.
Since the initial color electromagnetic fields ($\zeta=-1$) generate only non-zero z-component of the Lorentz force, $\kappa_L$ vanishes identically, while $\kappa_T$ is largest.
As $\zeta$ increases, $\kappa_L$ increases while $\kappa_T$ decreases.
In isotropic color fields ($\zeta=+1$), $\kappa_T$ is still larger than $\kappa_L$ though, as $\kappa_T$ receives an additional contribution from color magnetic fields, as seen in Eq.~(\ref{kappaf}).

\begin{figure}[h]
\centering
\includegraphics[width=0.5\textwidth]{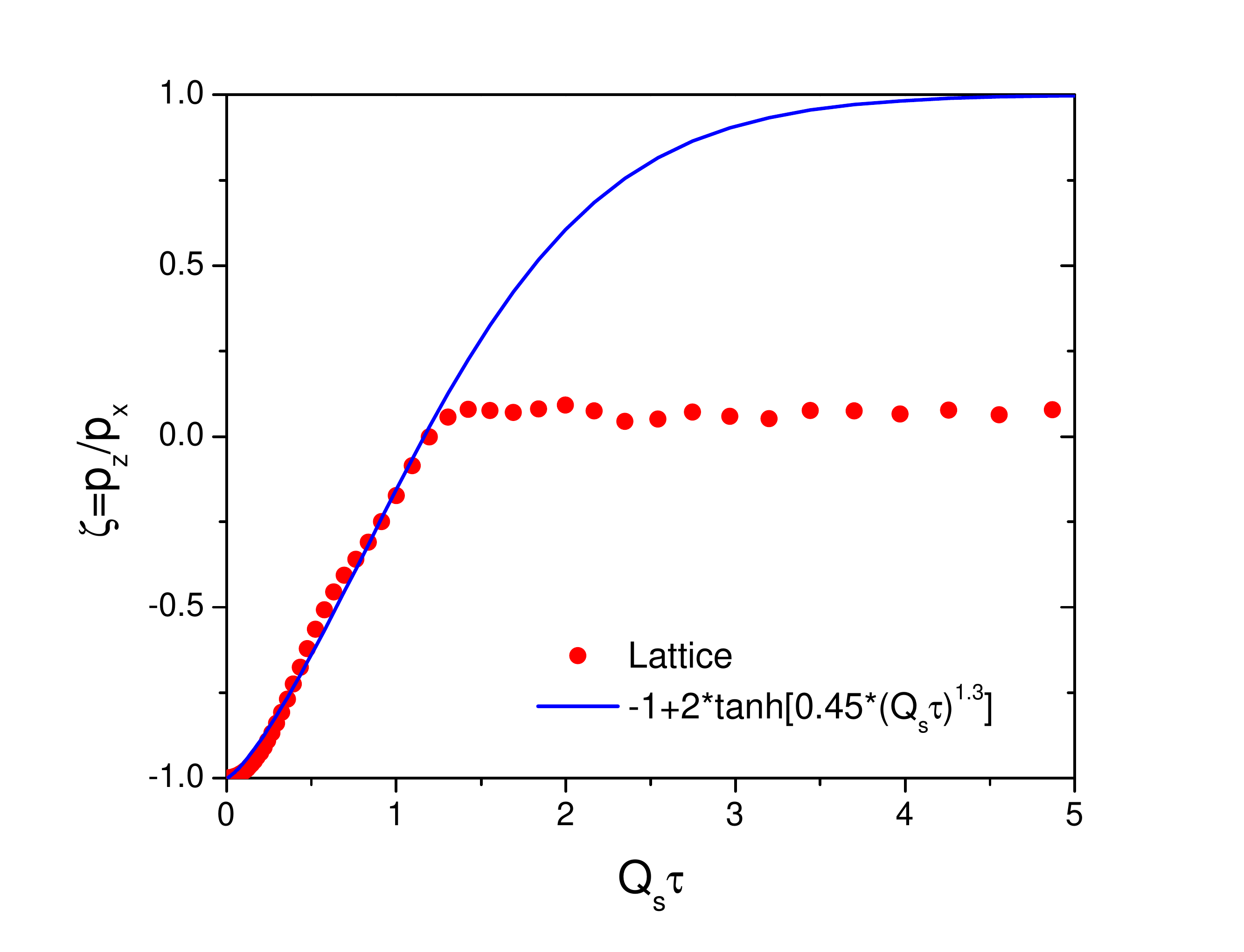}
\caption{(Color online) pressure anisotropy $\zeta$ as a function of $Q_s\tau$ from the lattice calculations of classical Yang-Mills equations~\cite{Gelis:2013rba} and from a parametrization.}
\label{anisotropy}
\end{figure}

Fig.~\ref{anisotropy} shows the pressure anisotropy $\zeta$ from the lattice calculations of classical Yang-Mills equations~\cite{Gelis:2013rba}. Computations were performed on a $512^2$ lattice with the transverse lattice spacing being fixed to $a_T=1$.
As mentioned before, $\zeta$ from the classical Yang-Mills equations starts from -1 and saturates around 0.
Assuming that complete solutions beyond those of classical Yang-Mills equations lead to $\zeta=+1$ at large $Q_s\tau$, we parameterize
\begin{eqnarray}
\zeta(\tau)=-1+2\tanh[0.45(Q_s\tau)^{1.3}],
\label{xitau}
\end{eqnarray}
which is shown as a blue line in Fig.~\ref{anisotropy}.
The parameterized function reproduces lattice results up to $Q_s\tau\simeq 1.2$ and converses to $\zeta=+1$.

The time-evolution of the energy density is obtained by using Bjorken's law
\begin{eqnarray}
\frac{\partial \varepsilon}{\partial \tau}=-\frac{\varepsilon +p_L}{\tau}=-\frac{2(\zeta +1)}{\zeta+2}\frac{\varepsilon}{\tau}
\end{eqnarray}
whose solution is
\begin{eqnarray}
\varepsilon(\tau)\sim \tau^{-\frac{2(\zeta +1)}{\zeta+2}}.
\end{eqnarray}

We note that the energy density is initially constant ($\zeta=-1$), then decreases like free streaming ($\zeta=0$) and finally isentropically ($\zeta=+1$).

\begin{figure}[h]
\centering
\includegraphics[width=0.5\textwidth]{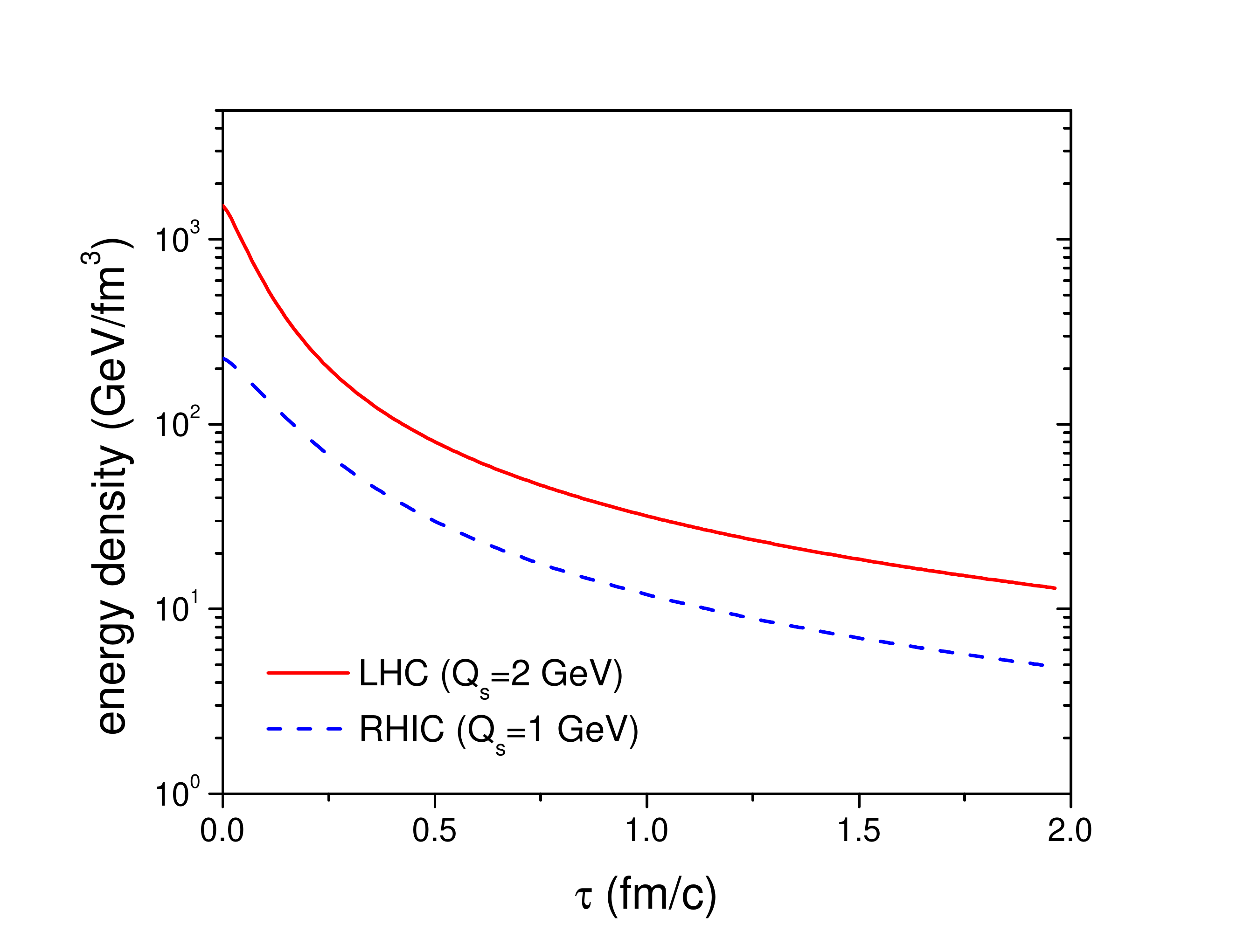}
\caption{(Color online) energy densities as a function of $\tau$ in central heavy-ion collisions at RHIC and LHC where $Q_s$ are respectively 1 and 2 GeV.}
\label{edn}
\end{figure}

Fig.~\ref{edn} shows the energy densities as a function of $\tau$ in central heavy-ion collisions at RHIC and LHC where $Q_s$ are taken to be respectively 1 and 2 GeV.
Initial energy densities $\varepsilon(\tau=0)$ are rescaled to meet the initial conditions of hydrodynamic simulations around $\tau=0.5$ fm/c~\cite{Song:2011kw}.

Then the diffusion coefficients of a heavy quark in relativistic heavy-ion collisions are given from Eq.~(\ref{kappa}) by

\begin{eqnarray}
\kappa_L(\tau,v)=\frac{2\pi\alpha_s}{N_c^2-1}\int_{\tau_0}^{\tau}d\tau^\prime\varepsilon(\tau^\prime)
\frac{\zeta(\tau^\prime)+1}{\zeta(\tau^\prime)+2}\nonumber\\
\times\exp\bigg[-\frac{v^2Q_s^2}{2}(\tau-\tau^\prime)^2\bigg],\\
\kappa_T(\tau,v)=\frac{\pi\alpha_s}{N_c^2-1}\int_{\tau_0}^{\tau}d\tau^\prime\varepsilon(\tau^\prime)
\frac{\zeta(\tau^\prime)+3}{\zeta(\tau^\prime)+2}\nonumber\\
\times(1+v^2)\exp\bigg[-\frac{v^2Q_s^2}{2}(\tau-\tau^\prime)^2\bigg].
\label{kappaf2}
\end{eqnarray}
where the heavy quark production time $\tau_0=1/(2M_T)$ with $M_T$ being the transverse mass of heavy quark.

\begin{figure}[h]
\centering
\includegraphics[width=0.5\textwidth]{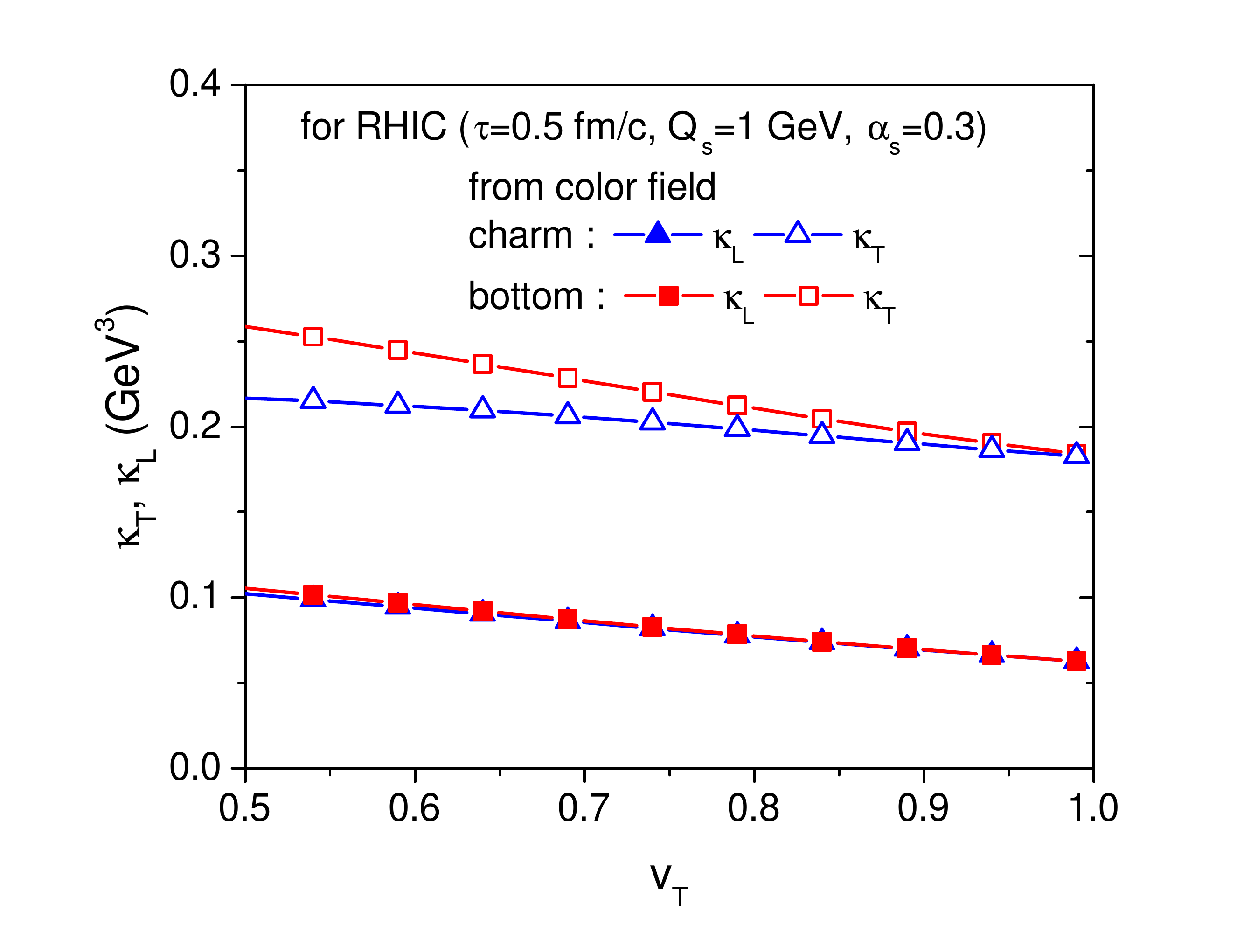}
\caption{(Color online) $\kappa_L$ (filled) and $\kappa_T$ (open) of a mid-rapidity charm (triangle) and bottom (square) quarks as a function of heavy-quark velocity at $\tau=0.5~{\rm fm/c}$ in central Au+Au collisions at $\sqrt{s_{\rm NN}}=$200 GeV. $Q_s$ and $\alpha_s$ are respectively taken to be 1 GeV and 0.3.}
\label{xi}
\end{figure}

Fig.~\ref{xi} shows $\kappa_L$ and $\kappa_T$ of a mid-rapidity heavy quark as a function of the heavy-quark velocity at $\tau=0.5$ fm/c in central Au+Au collisions at $\sqrt{s_{\rm NN}}=$200 GeV.
$Q_s$ and $\alpha_s$ are respectively taken to be 1 GeV and 0.3.
We can see that the $\kappa_T$ of bottom quark is larger than that of charm quark due to the smaller production time $\tau_0$ of bottom quark in Eq.~(\ref{kappaf2}).
Since the production times of both charm and bottom quarks become short for large velocity, the difference of $\kappa_T$ decreases with increasing heavy quark velocity.
On the other hand, the $\kappa_L$ of bottom quark is similar to that of charm quark regardless of velocity, because $\kappa_L$ is very small near the production time of heavy quark ($\zeta\approx -1$).

\section{Comparison with the diffusion coefficients induced by collisions}\label{collision}

As the system expands, the initial color fields become more and more dilute, and they can start to be described by individual quanta, that is, gluons.
In this section, we compare $\kappa_L$ and $\kappa_T$ of heavy quarks from coherent color fields with those induced by collisions within a thermalized quark-gluon plasma (QGP) at the same energy density.

Since the initial nuclear matter in relativistic heavy-ion collisions is highly occupied, the interspace between color quanta is short and perturbative QCD (pQCD) is applicable.
To the leading order in pQCD, $\kappa_L$ and $\kappa_T$ of heavy quark from collision read as followings~\cite{Moore:2004tg}:
\begin{eqnarray}
\kappa_L=\frac{4\pi C_H \alpha_s^2T^3}{3}\bigg(\frac{1}{v^2}-\frac{1-v^2}{2v^3}\ln\frac{1+v}{1-v}\bigg)~~~~~~~~~~~~~~~~~\nonumber\\
\times\bigg\{N_c[\ln(T/m_D)+C_b(v)]+\frac{N_f}{2}[\ln(T/m_D)+C_f(v)]\bigg\},\nonumber\\
\kappa_T=\frac{4\pi C_H \alpha_s^2T^3}{3}\bigg(\frac{3}{2}-\frac{1}{2v^2}+\frac{(1-v^2)^2}{4v^3}\ln\frac{1+v}{1-v}\bigg)~~~~~~\nonumber\\
\times\bigg\{N_c[\ln(T/m_D)+B_b(v)]+\frac{N_f}{2}[\ln(T/m_D)+B_f(v)]\bigg\},\nonumber\\
\label{kappac}
\end{eqnarray}
where $C_H=(N_c^2-1)/(2N_c)$, $N_f$ is the number of light flavors, $B_b(v),~B_f(v), C_b(v),$ and $C_f(v)$ are functions of the heavy quark velocity~\cite{Moore:2004tg}, and $m_D$ is the Debye screening mass, which is taken to be $1.5~T$ with $T$ being the temperature.

\begin{figure}[h]
\centering
\includegraphics[width=0.5\textwidth]{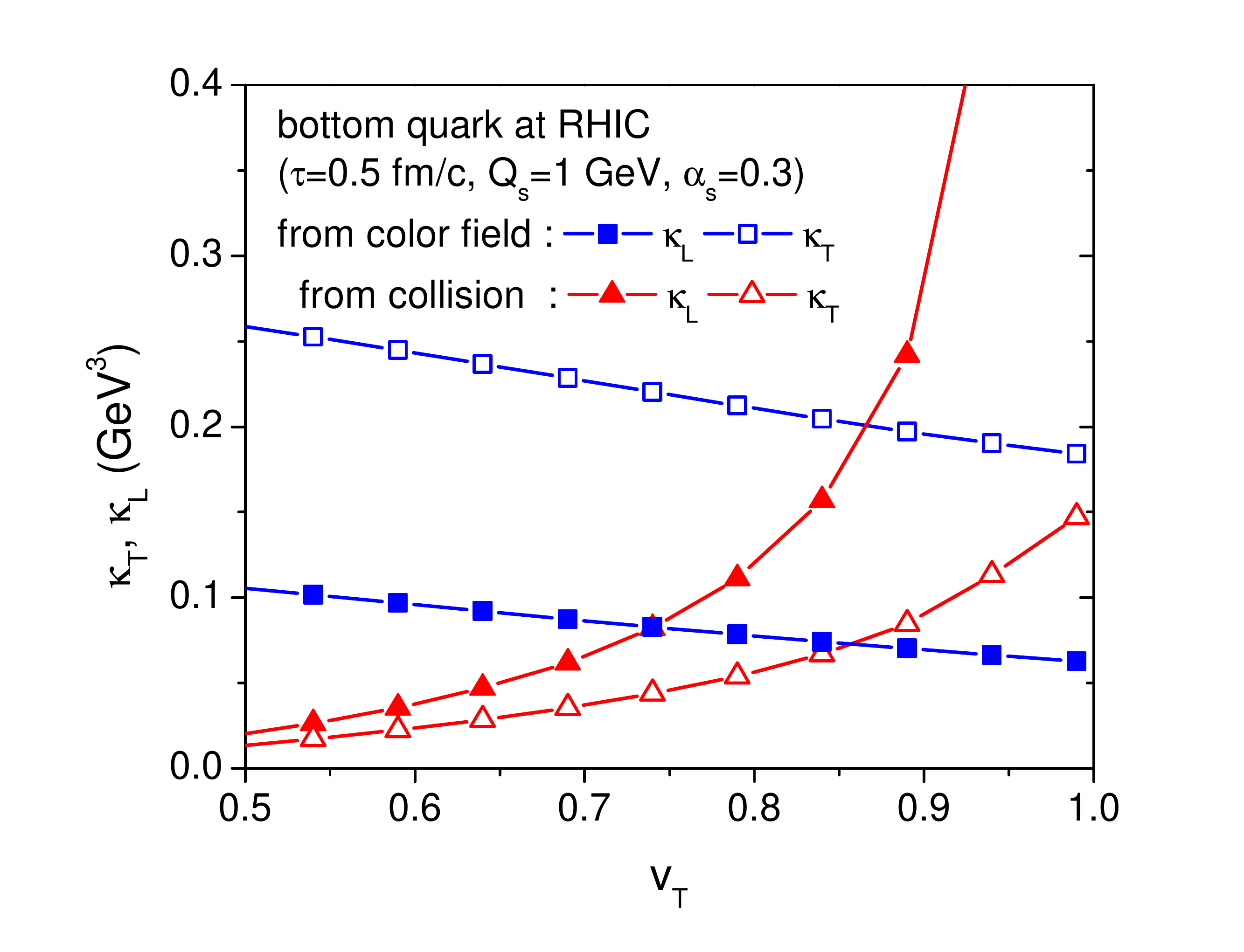}
\includegraphics[width=0.5\textwidth]{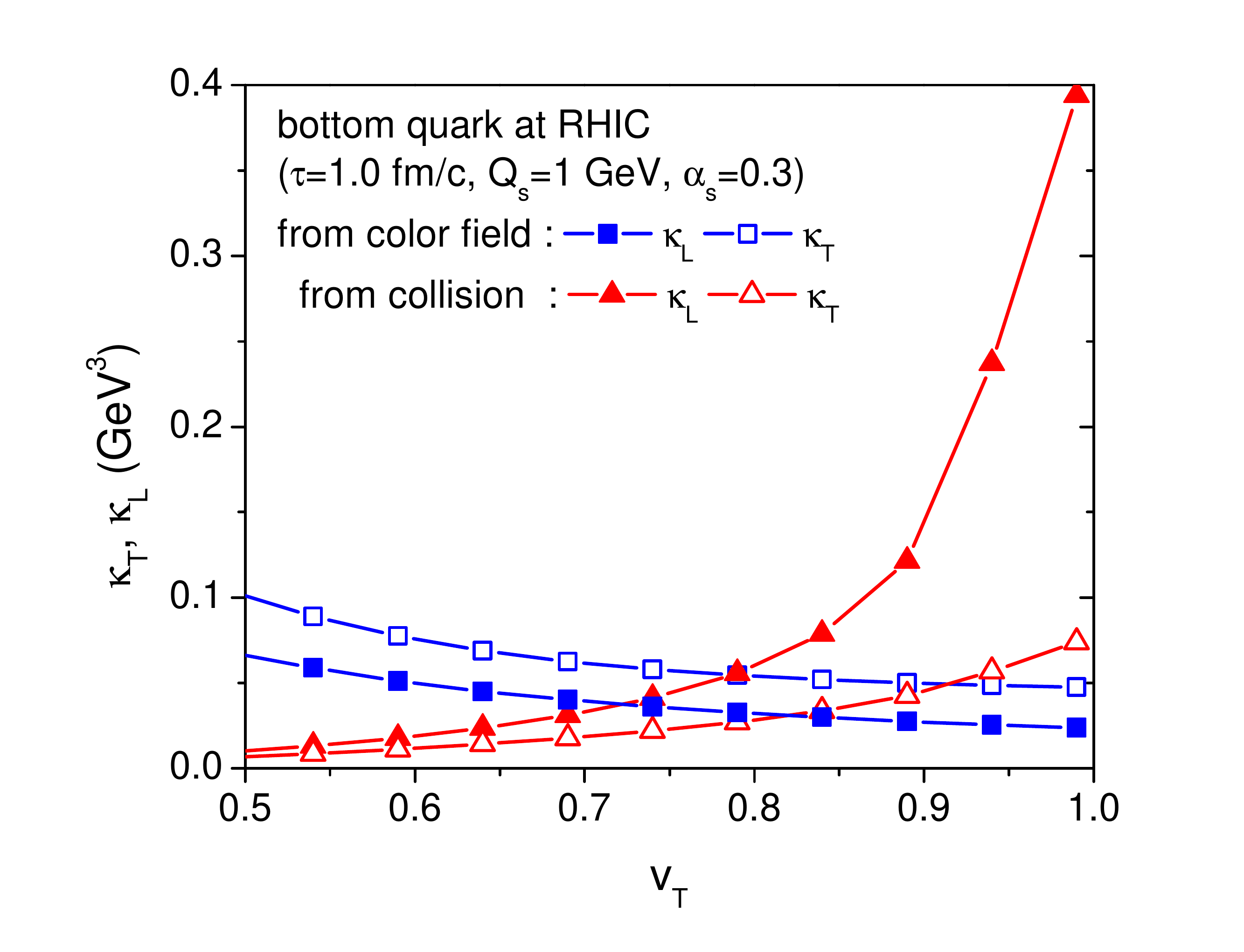}
\caption{(Color online) $\kappa_L$ (filled) and $\kappa_T$ (open) of a mid-rapidity bottom quark induced by collisions within a thermalized QGP (triangle), and from coherent color fields (square) at $\tau=0.5~{\rm fm/c}$ (upper) and $1.0~{\rm fm/c}$ (lower) in central Au+Au collisions at $\sqrt{s_{\rm NN}}=$200 GeV~\cite{Song:2011kw}.
The saturation momentum is taken to be 1.0 GeV, and $\alpha_s=$ 0.3 for both collisions and color Lorentz force.}
\label{rhic}
\end{figure}

In Fig.~\ref{rhic}, red lines with filled and open triangles show, respectively, $\kappa_L$ and $\kappa_T$ for collisions at $\tau=0.5$ fm/c and 1.0 fm/c in central Au+Au collisions at $\sqrt{s_{\rm NN}}=$200 GeV.
Temperatures in Eq.~(\ref{kappac}) are obtained from
\begin{eqnarray}
\varepsilon=\bigg(\frac{d_g}{30}+\frac{7d_q}{120}\bigg)\pi^2T^4,
\label{eden2}
\end{eqnarray}
where $d_g=16$ and $d_q=18$ are respectively the degrees of freedom of gluons and of light quarks.
Energy densities are 30 $\rm GeV/fm^3$ and 12 $\rm GeV/fm^3$, respectively, at $\tau=0.5$ fm/c and 1.0 fm/c from particle multiplicities in central Au+Au collisions at $\sqrt{s_{\rm NN}}=$200 GeV~\cite{Song:2011kw}.
They are compared with $\kappa_L$ and $\kappa_T$ from coherent color fields, which are shown as blue lines with filled and open squares.
The saturation momentum is taken to be 1.0 GeV, and $\alpha_s=$ 0.3 for both collisions and color Lorentz force.
Comparing the diffusion coefficients from color fields and those from collisions, we find two qualitative differences:
\begin{itemize}
\item[$\bullet$] Firstly, $\kappa_T$ from color fields is always larger than $\kappa_L$, whereas the opposite relation holds for collisions.
Since $\kappa_L$ is related to the drag coefficient and energy loss of energetic heavy quarks,
the coherent color fields are less effective for heavy-quark energy loss when compared to collisions.

\item[$\bullet$] Secondly, $\kappa_L$ and $\kappa_T$ from collisions increase with the heavy-quark velocity, while those from color fields decrease.
Since fast heavy quarks have more chance of colliding within the QGP and one collision has larger impact on it, $\kappa_L$ and $\kappa_T$ from collisions increase with heavy quark velocity.
On the other hand, fast heavy quarks remain only during a small amount of time within one color domain of the transverse plane.
It thus shortens the memory time defined in Eq~(\ref{memoryt}), and, as a result, $\kappa_L$ and $\kappa_T$ from color fields decrease.
\end{itemize}

In the upper panel of Fig.~\ref{rhic} where $\tau=0.5~{\rm fm/c}$, $\kappa_T$ from the color Lorentz force is larger than the one induced by collisions, and the same holds true for $\kappa_L$ up to $v=$ 0.74.
In the lower panel where $\tau=1.0~{\rm fm/c}$, however, $\kappa_T$ from the color Lorentz force is larger up to $v=$ 0.92, and $\kappa_L$ up to $v=$ 0.72.
This shows that as time increases, the diffusion coefficients from coherent color fields become less important, compared to those induced by collisions in relativistic heavy-ion collisions.
There are two reasons for that.
In Eq.~(\ref{kappaf2}), $\kappa_L$ and $\kappa_T$ from color fields are proportional to the energy density, while those from collisions to $T^3$ or $\varepsilon^{3/4}$ from Eq.~(\ref{kappac}) and (\ref{eden2}).
Furthermore, $\kappa_L$ and $\kappa_T$ from color fields are proportional to $\alpha_s$ while those induced by collisions to $\alpha_s^2$.
Therefore, coherent color fields are more important for heavy quark diffusion in the initial stage of relativistic heavy-ion collisions, where $\varepsilon$ is large and $\alpha_s$ is small.
If the saturation momentum decreases, then the transverse size of the color-field domains grows and the effect of coherent color fields becomes stronger for a longer time.
Comparing the upper and lower panels of Fig.~\ref{rhic}, we also find that $\kappa_L$ and $\kappa_T$ from color fields are early overtaken by those induced by collisions for fast heavy quarks, and late overtaken for slow heavy quarks.
Assuming that $\kappa_L$ and $\kappa_T$ from color fields turn into those induced by collisions as time increases, this suggests that initial coherent color fields are early seen as individual gluons by fast heavy quarks.
This seems reasonable, because the density of glasma decreases in the rest frame of fast heavy quarks.

\begin{figure}[h]
\centering
\includegraphics[width=0.5\textwidth]{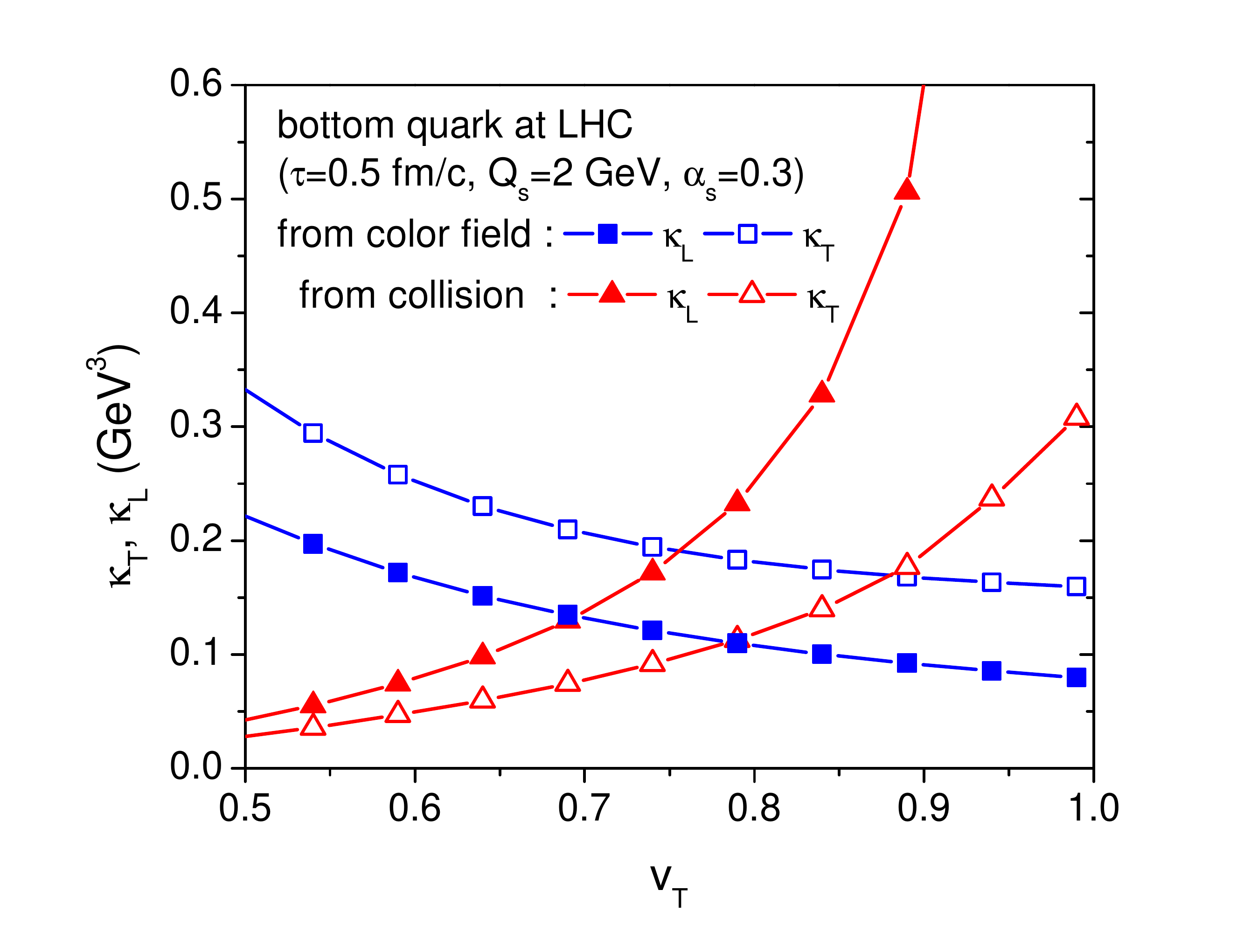}
\includegraphics[width=0.5\textwidth]{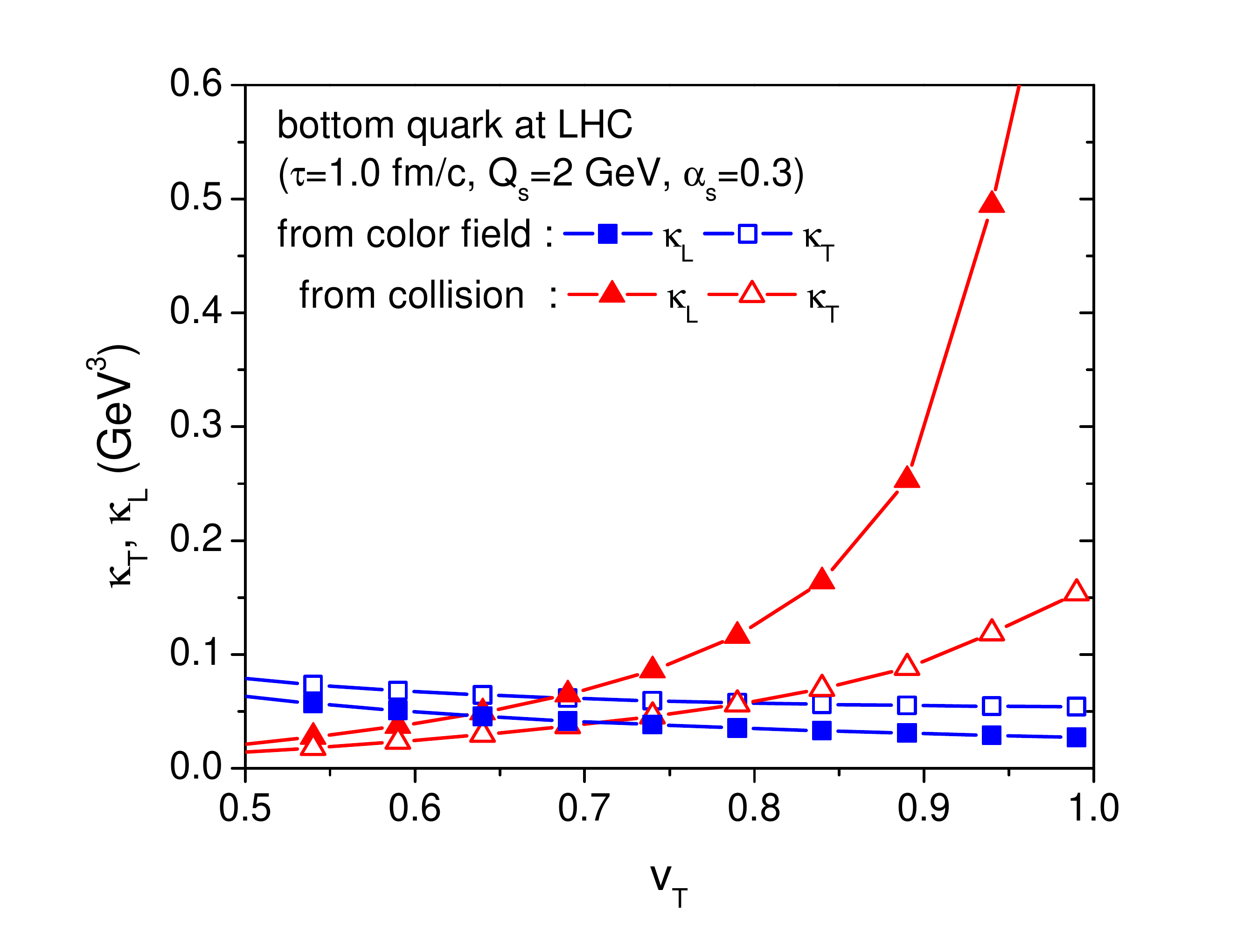}
\caption{(Color online) $\kappa_L$ (filled) and $\kappa_T$ (open) of a mid-rapidity bottom quark from collisions within thermalized QGP (triangle), and from coherent color fields (square) at $\tau=0.5~{\rm fm/c}$ (upper) and $1.0~{\rm fm/c}$ (lower) in central Pb+Pb collisions at $\sqrt{s_{\rm NN}}=$2.76 TeV~\cite{Song:2013tla}.
The saturation momentum and $\alpha_s$ are respectively taken to be 2.0 GeV and 0.3 in both panels.}
\label{lhc}
\end{figure}

Fig.~\ref{lhc} compares $\kappa_L$ and $\kappa_T$ from coherent color fields and those induced by collisions in central Pb+Pb collisions at 2.76 TeV.
The saturation momentum and $\alpha_s$ are respectively taken to be 2.0 GeV and 0.3.
$\kappa_T$ from color fields is larger than that from collision up to $v=$ 0.89 in the upper panel, and up to $v=$ 0.79 in the lower panel.
On the other hand, $\kappa_L$ from color fields is larger than the one induced by collisions up to $v=$ 0.69 and $v=$ 0.64 in the upper and lower panels respectively.
Though energy densities in Pb+Pb collisions at $\sqrt{s_{\rm NN}}=$2.76 TeV are higher than in Au+Au collisions at $\sqrt{s_{\rm NN}}=$200 GeV, coherent color fields play a less important roles in heavy quark diffusion due to a larger saturation momentum.

\section{summary}\label{summary}

Heavy quark is a promising probe for the properties of the hot dense nuclear matter created in relativistic heavy-ion collisions.
Though the interactions of heavy quarks with thermalized hot dense nuclear matter have been extensively studied, the interactions before thermal equilibrium has been reached in relativistic heavy-ion collisions have barely been studied.
According to the CGC model, the initial nuclear matter is composed of strong coherent color fields, which are initially highly anisotropic and then presumably evolve towards an isotropic gluon gas.

In this study we have calculated the diffusion coefficients of heavy quarks induced by interactions with coherent color fields.
Due to our lack of knowledge of the correlation length of the color fields in $\tau$ and $\eta$ directions, our calculations were restricted to mid-rapidity heavy quarks with rather large velocities.
The coherent color fields were modeled such that they initially follow the numerical results from lattice calculations and then finally become isotropic.

We then compared the diffusion coefficients with those obtained from collisions within a thermalized QGP in Au+Au collisions at $\sqrt{s_{\rm NN}}=$200 GeV, and in Pb+Pb collisions at $\sqrt{s_{\rm NN}}=$2.76 TeV.
From this comparison, we have found a couple of qualitative differences between diffusion coefficients from color fields and those from collisions:
\begin{itemize}
\item[$\bullet$]Firstly, $\kappa_T$ calculated from interactions with color fields is always larger than $\kappa_L$ for a mid-rapidity heavy quark in relativistic heavy-ion collisions, which is opposite to the conclusion one draws from collisional interactions.
Therefore, coherent color fields are less effective for heavy-quark energy loss than collisions.

\item[$\bullet$]Secondly, $\kappa_L$ and $\kappa_T$ from color fields decrease with increasing heavy quark velocity, while those from collisions increase.
The reason for decreasing diffusion coefficients in color fields is that the time during which a heavy quark remains in one domain of color field decreases with increasing heavy quark velocity, and the short stay reduces the correlation of the color Lorentz forces.
\end{itemize}
We have also found that the contribution from coherent color fields to heavy quark diffusion is important in the early stage of relativistic heavy-ion collisions, because $\kappa_L$ and $\kappa_T$ from color fields are proportional to $\varepsilon$ and $\alpha_s$, while those from collisions vary as $\varepsilon^{3/4}$ and $\alpha_s^2$.

As our study is restricted to fast mid-rapidity heavy quarks, it would be very interesting to extend to heavy quarks with small velocity or in forward and backward rapidities. For this one would a priori just need the correlation length of coherent color fields in the $\tau$ and $\eta$ directions.

\section*{Acknowledgements}

This work has been supported by DFG under contract BR 4000/3-1, by BMBF under project No. 05P12RFFCQ, by the LOEWE center "HIC for FAIR", and by the Natural Sciences and
Engineering Research Council of Canada (NSERC). The computational resources have been provided by the LOEWE-CSC, as well as the Guillimin supercomputer at McGill
University, managed by Calcul Qu\'ebec and Compute Canada. The
operation of this supercomputer is funded by the Canada Foundation for
Innovation (CFI), Minist\`ere de l'Economie, de l'Innovation et des
Exportations du Qu\'ebec (MEIE), RMGA and the Fonds de recherche du
Qu\'ebec - Nature et technologies (FRQ-NT).

\section*{Appendix A: electric and magnetic field equality for purely gluonic systems}

To the best of our knowledge, the equality deduced at the end of this appendix has yet to be written somewhere. For completeness, let us roughly sketch its derivation. Starting from the Classical Yang-Mills equations written in a covariant form
\begin{align}
\frac{1}{\sqrt{-g}}D_\mu^{ab}\left[\sqrt{-g}g^{\mu\alpha}g^{\nu\beta}F^b_{\alpha\beta}\right]
=\null&0\;,\label{eq:ymeom}
\end{align}
where $g={\rm det}g_{\mu\nu}$, the covariant derivative being defined from the gluonic fields as
\begin{align}
D_\mu^{ab}=\delta^{ab}\partial_\mu-igA_\mu^{ab}\;,
\end{align}
and the Maxwell-stress tensor being
\begin{align}
F_{\mu\nu}^{a}=\null&\partial_\mu A_\nu^{a}-\partial_\nu A_\mu^{a}-ig[A_\mu,A_\nu]^a\;.
\end{align}
Placing ourselves in the Milne coordinate system
\begin{align}
\tau=\null&\sqrt{t^2-z^2}&\eta=\null\frac12\ln\frac{t+z}{t-z}\;,
\end{align}
where the metric tensor reads
\begin{align}
g_{\mu\nu}=\null& {\rm diag}(1,-1,-1,-\tau^2)\label{eq:metric}
\end{align}
and defining as is usually done in this coordinate system the electric fields as (here $i$ runs on only on the transverse spatial coordinates $x,y$)
\begin{align}
E^{ia}=\null&\tau\partial_\tau A_i^a&
E^{\eta a}=\null&\frac{1}{\tau}\partial_\tau A_\eta^a\;,
\end{align}
we get a constraint equation (Gauss's law) by picking $\nu=\tau$ in Eq. (\ref{eq:ymeom})
\begin{align}
D_iE^i+D_\eta E^\eta=\null&0\;,\label{eq:gausslaw}
\end{align}
and three equations for the time derivative of the transverse and longitudinal electric fields by picking$\nu=i,\eta$ in Eq. (\ref{eq:ymeom})
\begin{align}
\partial_\tau E^{ia}=\null&\tau D_j^{ab}F^b_{ji}+\frac{1}{\tau} D_\eta^{ab}F^b_{\eta i}\;,&
\partial_\tau E^{\eta a}=\null&\frac{1}{\tau} D_i^{ab}F^b_{i\eta}\;.\label{eq:ymel}
\end{align}
Deriving the constraint equation (\ref{eq:gausslaw}) trivially leads to
\begin{align}
D^{ab}_i\partial_\tau E^{ib}+D^{ab}_\eta\partial_\tau E^{\eta b}
-\frac{ig}{\tau}E^{ia}E^{ia}-ig\tau E^{\eta a}E^{\eta a}=\null&0\;,\label{eq:ymint1}
\end{align}
and using the following relation between the covariant derivative and the Maxwell-stress tensor
\begin{align}
[D_\mu,D_\nu]^a=\null& -ig F^a_{\mu\nu}\;,
\end{align}
we get, by plugging Eq. (\ref{eq:ymel}) into Eq. (\ref{eq:ymint1})
\begin{align}
\tau F_{xy}^{a} F_{xy}^{a}+\frac{1}{\tau} F_{i\eta}^{a} F_{i\eta}^{a}
-\frac{1}{\tau}E^{ia}E^{ia}-\tau E^{\eta a}E^{\eta a}=\null&0\;.
\end{align}
Finally, defining as usual the magnetic fields as
\begin{align}
B^{\eta a}=\null&F_{xy}^{a}\;,&
B^{x a}=\null&F_{y \eta}^{a}\;,&
B^{y a}=\null&F_{\eta x}^{a}\;,
\end{align}
we get
\begin{align}
\tau B^{\eta a}B^{\eta a} +\frac{1}{\tau} B^{i a}B^{i a}
=\null&\frac{1}{\tau}E^{ia}E^{ia}+\tau E^{\eta a}E^{\eta a}\;.
\end{align}
Given the metric form in  Milne coordinate system (\ref{eq:metric}), one has
\begin{align}
-E^a_\mu E^{\mu a}=E^{ia}E^{ia}+\tau^2 E^{\eta a}E^{\eta a}\;.
\end{align}
we therefore just proved that at all times (denoting the spatial indices $I=x,y,\eta$), the following equality relates the electric fields and the magnetic fields
\begin{align}
E^a_I E^{I a}=\null& B^a_I B^{I a}\;.
\end{align}

\section*{Appendix B: electric and magnetic field relations in the free-streaming case}

Solving numerically the classical Yang-Mills equations~\cite{Lappi:2006fp}, one finds that after some time
\begin{align}
p_x\approx p_y \approx \null&\frac{\epsilon}{2}&
p_z\approx\null& 0\;,
\label{eq:pressures}
\end{align}
where, calling
\begin{align}
E_x^2=\null&\frac{E_x^aE_x^a}{\tau^2}&
E_y^2=\null&\frac{E_y^aE_y^a}{\tau^2}&
E_z^2=\null&E_\eta^aE_\eta^a\;,\notag\\
B_x^2=\null&\frac{B_x^aB_x^a}{\tau^2}&
B_y^2=\null&\frac{B_y^aB_y^a}{\tau^2}&
B_z^2=\null&B_\eta^aB_\eta^a\;,
\end{align}
one has
\begin{align}
\epsilon=\null&\frac12\left(E^2+B^2\right)\;,&
p_i=\null&\epsilon-E_i^2-B_i^2\;.
\end{align}
Given what we proved in the previous appendix, one has the additional relation valid at all times
\begin{align}
E_x^2+E_y^2+E_z^2=\null&B_x^2+B_y^2+B_z^2\;.\label{eq:c1}
\end{align}
Equation  (\ref{eq:pressures}) gives only two additional constraints on the fields  :
\begin{align}
E_z^2+B_z^2=\null&E_x^2+B_x^2+E_y^2+B_y^2\;,\label{eq:c2}
\end{align}
and
\begin{align}
E_x^2+B_x^2=E_y^2+B_y^2\;.\label{eq:c3}
\end{align}
Combining Eq. (\ref{eq:c1}) with Eq. (\ref{eq:c2})-(\ref{eq:c3}) leads to the following possible rewritting of the constraints
\begin{align}
E_z^2=\null&B_x^2+B_y^2&
B_z^2=\null&E_x^2+E_y^2\;.
\end{align}
Making the additional assumptions (that seems reasonable after some time evolution, looking the numerical results of~\cite{Lappi:2006fp}) that
\begin{eqnarray}
E_x^2+E_y^2&=&B_x^2+B_y^2\nonumber\\
E_z^2&=&B_z^2,
\label{aniso12}
\end{eqnarray}
we get
\begin{eqnarray}
E_z^2 = E_x^2+E_y^2,\nonumber\\
B_z^2 = B_x^2+B_y^2.
\end{eqnarray}

\end{document}